\documentclass{article}

\usepackage{arxiv}

\usepackage[utf8]{inputenc} 
\usepackage[T1]{fontenc}    
\usepackage{hyperref}       
\usepackage{url}            
\usepackage{booktabs}       
\usepackage{amsfonts}       
\usepackage{nicefrac}       
\usepackage{microtype}      
\usepackage{graphicx}
\usepackage{amsmath}
\usepackage{tikz}
\usetikzlibrary{arrows.meta,positioning,shapes.geometric}
\usepackage{xcolor}
\usepackage[numbers]{natbib}

\title{Plans They Abandon, Reports They Author: The Narrative Layer of Autonomous Agents}

\author{
  Obada Kraishan \\
  College of Media and Communication \\
  Texas Tech University \\
  Lubbock, TX 79409, USA \\
  \texttt{omareikr@ttu.edu} \\
  \And
  Kulsawasd Jitkajornwanich \\
  College of Media and Communication \\
  Texas Tech University \\
  Lubbock, TX 79409, USA \\
  \texttt{kulsawasd.jitkajornwanich@ttu.edu} \\
}

\begin{document}
\maketitle

\begin{abstract}
When a coding agent finishes a task, the developer reviews a summary the agent
wrote about itself, not a display someone designed. We ask how much of the
agent's work that summary carries, and whether it drifts toward the plan the
agent stated when execution departed from it. Across 5,851 real developer
sessions and 355,942 tool calls, a self-report referred to about one action in
eleven, and a reader working from the report alone recovered roughly a fifth of
the action log. Neither figure depended on whether the session later needed
human correction. Reports did not generally resemble the stated plan more than
the executed one, but they did so increasingly as execution diverged from the
plan. We hand-validate both measurement steps that use a language model, report
the one that failed alongside the one that passed, and draw conclusions only
from measures that survived.
\end{abstract}


\section{Introduction}

A developer delegates a task to a coding agent, steps away, and returns to a message. The message says what was done: files edited, tests run, a bug fixed. The developer reads it, perhaps glances at a diff, and moves on. This is now an ordinary working pattern, and it rests on an assumption that the HCI literature on trust in automation has never had to examine: that the account of the system's behaviour was written by the system itself.

Trust calibration theory locates appropriate reliance in the quality of the information the operator receives about the automation \cite{lee2004trust, parasuraman2000model}. In that literature the information is designed. A display shows confidence; a log records events; an explanation is produced by a mechanism the designer chose. Language-model agents break this arrangement. The agent's plan, its running commentary, and its completion summary are all generated text, produced by the same process whose behaviour they describe. Studies of how people oversee such agents report that engagement with the agent's output declines over the course of a task \cite{notreading2026}, that action traces in their current form are cumbersome to verify against \cite{grunde2026overseeing}, and that developers lean on summaries and spot checks rather than full traces \cite{dhanorkar2026oversight}. Whether the summaries are accurate is a separate question, and it is the one this paper takes up.

We approach the agent's text as a \emph{narrative layer} with two halves. Prospectively, agents state plans. Retrospectively, they author reports. Both can be compared against a third record that is not narrated at all: the log of tool calls the agent actually executed. Suchman argued four decades ago that plans are resources for action rather than controllers of it, and that the appearance of plan-following is largely reconstructed after the fact \cite{suchman1987plans, suchman1993response}. Earlier planning systems left no narrative behind, so the claim resisted quantitative test. Agents that write their plans down, act, and then write down what they did make it testable at scale.

This paper reports such a test on 5,851 real developer sessions from the SWE-chat corpus \cite{swechat2026}. We ask two questions.

\begin{itemize}
  \item \textbf{RQ1.} How much of what an agent does appears in its self-report?
  \item \textbf{RQ2.} Does the self-report drift toward the stated plan as execution diverges from it?
\end{itemize}

For RQ1 we measure omission, the share of executed actions no claim in the report refers to, and recoverability, the degree to which a reader given only the report could reconstruct the log. For RQ2 we define a repair index that is positive when a report resembles the plan more than it resembles execution, and we test whether it rises with plan-execution divergence.

Our contributions are empirical and methodological. Empirically, we show that self-reports carry a small share of what an agent did, that this share holds whether or not the session later required human correction, and that narrative repair is conditional rather than general: reports track execution when agents follow their plans and drift toward the plan when they do not. Methodologically, we show how the three records an agent leaves behind can be aligned and scored on a shared vocabulary, we validate both model-dependent measurement steps against hand coding, and we report the one that failed alongside the one that passed. The analysis pipeline, the prompts, and the derived data are available from the authors on request and will be released publicly.

\section{Related Work}

This section situates the paper against three bodies of work: trust and transparency in automation, agent traces as objects of study, and work that addresses agent self-reporting directly.

\subsection{Transparency information and its author}

Lee and See's account of trust in automation distinguishes trust from trustworthiness and locates calibration in the operator's access to accurate information about the system's performance, process, and purpose \cite{lee2004trust}. Parasuraman, Sheridan, and Wickens' levels-of-automation framework likewise assumes that the human's picture of the automation is mediated by displays the designer specified \cite{parasuraman2000model}. Later HCI work on AI-assisted decision-making inherits this framing: explanations are treated as designed artefacts whose form the researcher can vary and whose effects on reliance can be measured \cite{bansal2021whole, bucinca2021trust, amershi2019guidelines}.

Generative agents complicate this picture because the explanation and the behaviour share a source. Tankelevitch et al. describe the metacognitive demands that generative systems place on users, including the need to evaluate outputs one did not produce \cite{tankelevitch2024metacognitive}. Delegation to an agent extends that demand from evaluating an output to evaluating an account of a process, and the account is itself an output. Chan et al. argue for visibility into agent behaviour through mechanisms external to the agent, such as identifiers, activity logs, and real-time monitoring, precisely because self-description cannot be assumed reliable \cite{chan2024visibility}. Work on chain-of-thought reasoning has found that models' stated reasoning can diverge from the process that produced their answers \cite{turpin2023unfaithful, lanham2023faithfulness}. We extend that concern from a single answer to a long-horizon task with a plan at one end and a summary at the other.

\subsection{Agent traces and how they are read}

Agent scaffolds interleave reasoning and action \cite{yao2023react}, and many add reflective summaries at the end of a task or turn \cite{shinn2023reflexion}. Software-engineering agents in particular emit a rich trace: tool calls with arguments, results, and interleaved natural-language commentary \cite{yang2024sweagent, jimenez2024swebench}. Surveys of the area treat the trace as an engineering artefact for debugging and evaluation \cite{wang2024survey}.

HCI has begun to ask how people actually engage with these traces. Grunde-McLaughlin et al. found in three user studies that basic action traces were cumbersome for verification, and that an interface designed to help participants find errors improved their confidence more than their accuracy \cite{grunde2026overseeing}. A formative study with four software engineers observed cognitive engagement declining as a task progressed and participants overlooking details of what the assistant produced \cite{notreading2026}. Interviews with seventeen developers describe oversight in practice as spot-checking and heuristic judgement rather than full review \cite{dhanorkar2026oversight}, and a comparison of oversight strategies for computer-use agents found that strategy shaped users' exposure to problematic actions more than their ability to correct them \cite{oversight2026}. Our question is prior to all of these: whatever the reader does with the summary, what is in it?

\subsection{Checking what agents claim}

A growing body of work addresses the gap between what agents claim and what they did. Tang et al. analysed 20,574 real coding-agent sessions and identified seven recurring forms of developer-agent misalignment, one of which concerns how agents report progress; most episodes imposed effort and trust costs rather than damage, and most visible resolutions required explicit user correction \cite{codingagentsfail2026}. Systems work has proposed claim-to-evidence trace graphs for auditing agent sessions \cite{ledger2026}, artefact-centred observability for scientific agents \cite{claimaware2026}, interactive evidence extraction from web-agent trajectories \cite{hansel2026}, and generation-time provenance records that tie each claim to the tool observation supporting it \cite{tracer2026}. These are tools: they aim to solve the problem by making claims checkable.

This paper is complementary. It does not build a checker. It measures the phenomenon on in-the-wild sessions, characterises its structure, and connects it to two theoretical accounts, trust calibration and situated action, that predict where and why self-reports should depart from execution. The measurement is also two-sided: prior work concentrates on the retrospective report, whereas we analyse the prospective plan alongside it and test how the two relate.

\subsection{Language models as measurement instruments}

Two of our measurement steps read text with a language model, so the literature on model-based measurement bears on whether the results can be believed. Judges show position and self-preference biases \cite{zheng2023judging, panickssery2024selfpreference}, output is sensitive to prompt formatting \cite{sclar2024quantifying}, behaviour drifts between versions of the same endpoint \cite{chen2024changing}, and temperature-zero decoding is not deterministic in practice \cite{he2025nondeterminism}. Extraction carries a risk of its own: a model asked to list the intentions stated in a passage can return intentions the passage does not contain.

Pang et al. find that CHI papers using language models frequently name such concerns without acting on them \cite{pang2025llmification}. Section~\ref{sec:validation} reports what we did about each. Prompts are fixed and released verbatim, and every call names a dated snapshot rather than an alias. Both model-dependent steps were checked against hand coding; one check passed and one did not, and the analysis is built accordingly. Nondeterminism we did not address, and list among the limitations. Following Agnew et al.'s caution about participant-free research that makes claims about people \cite{agnew2024illusion}, we note that every claim here concerns the behaviour of a software system.

\section{The Narrative Layer}

\begin{figure}[t]
  \centering
  \begin{tikzpicture}[
      node distance=9mm and 16mm,
      rec/.style={draw, rounded corners=2pt, thick, minimum width=30mm, minimum height=11mm, align=center, font=\small},
      meas/.style={draw, dashed, rounded corners=2pt, minimum width=24mm, align=center, font=\footnotesize, fill=gray!8},
      arr/.style={-{Latex[length=2.5mm]}, thick}]
    \node[rec, fill=blue!7]   (plan)   {Stated plan\\ \footnotesize prospective narration};
    \node[rec, fill=orange!8, right=of plan]   (log)    {Executed log\\ \footnotesize tool calls, not narrated};
    \node[rec, fill=blue!7,  right=of log]    (report) {Self-report\\ \footnotesize retrospective narration};
    \draw[arr] (plan) -- (log);
    \draw[arr] (log) -- (report);
    \node[meas, below=of log, yshift=2mm] (div) {\textbf{Divergence}\\ $1 - \mathrm{LCS}(\text{plan},\text{log}) / |\text{plan}|$};
    \node[meas, below=of report, yshift=2mm] (om) {\textbf{Omission}\\ share of log no claim references};
    \node[meas, below=of div, yshift=1mm, xshift=24mm] (rep) {\textbf{Repair}\\ $C(\text{report},\text{plan}) - C(\text{report},\text{log})$};
    \draw[arr, gray] (plan.south) to[out=-90,in=180] (div.west);
    \draw[arr, gray] (log.south)  -- (div.north);
    \draw[arr, gray] (log.south east) to[out=-60,in=170] (om.west);
    \draw[arr, gray] (report.south) -- (om.north);
    \draw[arr, gray] (div.south) to[out=-90,in=180] (rep.west);
    \draw[arr, gray] (om.south)  to[out=-90,in=0]   (rep.east);
  \end{tikzpicture}
  \caption{The three records an agent leaves behind and the three measures computed across them. The plan and the report are text the agent wrote; the log is the sequence of tool calls it executed. All three are expressed in a shared vocabulary of six action types before comparison.}
  \label{fig:layer}
\end{figure}
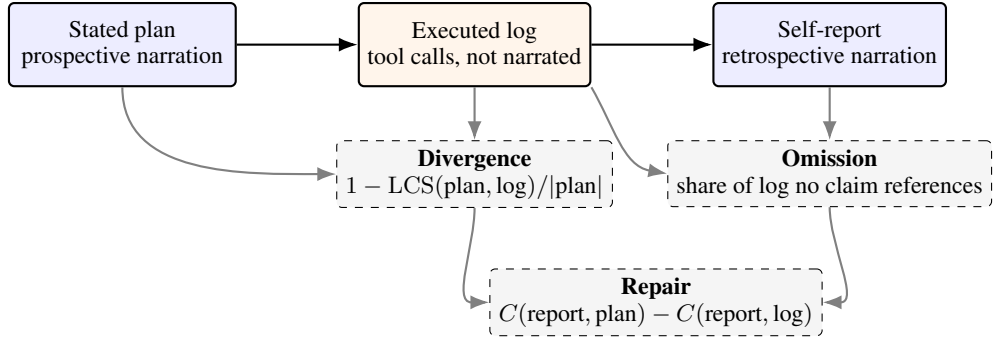

The object of this study is the text an agent produces about its own work. This section sets out what that text can be compared against, and why the comparison is possible at all. Figure~\ref{fig:layer} gives the overview; the measures it names are defined in Section~\ref{sec:measures}.

Every session leaves three records. The \emph{stated plan} is what the agent says it will do, either in prose (``I'll read the config, then update the handler, then run the tests'') or through a structured task-list tool that some scaffolds provide. The \emph{executed log} is the sequence of tool calls the agent made: reads, edits, shell commands, and so on, each with a target. The \emph{self-report} is the agent's final message to the user, in which it describes what it did and what resulted.

The plan and the report are narration. The log is not; it is recorded by the scaffold as calls are made, and the agent does not compose it. This asymmetry is what makes the comparison informative. Both narrations can be checked against a record neither of them wrote.

We call these two narrations together the \emph{narrative layer}. It is not a
by-product of the scaffold in the way the log is. It is written, it is what a
reader is given, and it is the only part of the record whose relation to the
work is an open question. The rest of the paper measures that relation.

\section{Method}

This section describes the corpus, the pipeline that produces the three sequences per session, the matching procedure behind omission, and the statistical approach. Every step is a standalone script in the released repository, and every result file records the commit, configuration hash, random seed, and model snapshot that produced it.

\subsection{Corpus}

We analyse SWE-chat \cite{swechat2026}, a public corpus of coding-agent sessions recorded from real developers working in their own repositories through a session-capture CLI. The release we used (revision \texttt{f66cca95}) contains 5,851 sessions, 2,692,480 transcript turns, 14,459 commits, and 205 repositories. Sessions come from several scaffolds; Claude Code accounts for 82.9\%, OpenCode for 10.6\%, Codex for 3.6\%, and Gemini CLI for 1.0\%, with the remainder spread across smaller tools. Fifteen distinct model versions appear. The corpus is released under ODC-BY-1.0.

Each transcript turn is typed. The types that matter here are \texttt{tool\_use} (355,942 turns), which give the executed log; \texttt{assistant\_response} (41,622) and \texttt{assistant\_thinking} (11,954), which carry the narration; and two structured task-list tools, \texttt{TaskCreate} and \texttt{TaskUpdate} (24,066 calls across 1,211 sessions), which give a machine-readable plan where the agent used one. Tool calls arrive pre-parsed with tool name, file path, and command, so canonicalisation to the six action types is a deterministic mapping rather than a per-scaffold parser. Of the 355,942 actions, 1,148 (0.3\%) matched a pattern for irreversible commands (forced pushes, hard resets, recursive deletes).

We considered two additional corpora for a non-coding replication and rejected both after inspecting their records. GUI-Odyssey stores each step as an action type and screen coordinates with no agent prose, and the AgentNet computer-use corpus was not processed within the study period. We return to this in Section~\ref{sec:limits}; it means every claim here concerns coding agents.

\subsection{A shared vocabulary of action types}

Direct comparison of prose with tool calls is not meaningful, so we map all three records onto a common set of six action types: \textsc{read} (inspecting files or searching), \textsc{write} (creating or editing), \textsc{execute} (running commands, tests, or builds), \textsc{network} (fetching remote resources), \textsc{orchestrate} (task-list and sub-agent operations), and \textsc{other}. Tool calls map by tool name and command. Plan units and claim units map by lexical cues in their text. A session then yields three sequences of action types, one per record, and the measures below are computed over those sequences.

\subsection{Segmenting the narration}

Assistant turns are labelled prospective or retrospective by position, without a model. A turn is \emph{prospective} if it is an \texttt{assistant\_thinking} turn, or if it is an \texttt{assistant\_response} that precedes the session's last tool call and contains plan language (an enumerated list, or a phrase such as ``I'll'', ``let me'', or ``next,''). The final \texttt{assistant\_response} of a session is the \emph{self-report}. This yielded 30,126 prospective spans and 5,450 self-reports.

Thinking turns are sparse in this corpus: only 670 sessions (11.5\%) contain one, and their median length is 40 characters, consistent with summarised rather than verbatim reasoning. Prose plans in response turns are the main prospective source.

\subsection{Extracting plan units and claim units}

Plan units and claim units were extracted with a pinned language-model snapshot (\texttt{claude-haiku-4-5-20251001}, temperature 0, one sample per item), using two fixed prompts released with the code. The plan prompt asks for the atomic actions the agent says it intends to take, as a JSON list in stated order, and returns an empty list when the passage states no intention. The claim prompt asks for assertions about what the agent did, did not do, or achieved, each typed as \textsc{action}, \textsc{non-action}, or \textsc{outcome} and flagged as verifiable when it names a file, command, or test result. Of 30,056 prospective spans, 27.4\% yielded at least one plan unit; the remaining spans were responses that stated no intention, for which an empty list is the correct output. Five spans in 35,411 returned unparseable output.

Structured plan units were taken directly from \texttt{TaskCreate} and \texttt{TaskUpdate} payloads with no model involved. Plan units from all sources were pooled per session and tagged with their provenance.

Pooling is a decision worth stating, because the two sources are not interchangeable. On the 535 sessions carrying both, structured and prose plan units agreed at a per-session $F_1$ of only 0.275 ($SD = 0.228$). A task list is typically written early and enumerates the work at the grain of the user's request; a prose plan typically appears later, after the agent has read code, and enumerates the next few steps at the grain of the code. Taking either source alone as ``the plan'' would measure divergence against one horizon and not the other, so we use both and note the choice as a limitation.

The pipeline produced 45,206 plan units (16,946 structured, 28,260 prose) and 20,641 claim units (12,831 \textsc{action}, 6,561 \textsc{outcome}, 1,075 \textsc{non-action}, 5 misclassified as plans and discarded).

\subsection{Aligning claims to actions}
\label{sec:alignment}

Omission requires a match from claims to log entries. Matching is rule-first. A claim that names a file path or command is matched lexically to the first log entry whose target contains that string. A \textsc{non-action} claim (``I did not modify the config'') inverts: it is supported when no log entry touches the named target. Claims the rules could not place were sent to a model adjudicator with the session's candidate actions and asked for a verdict. In the final data, 45.2\% of claim verdicts were decided by rules and 52.8\% by the adjudicator.

Omission uses only the \emph{matched action indices}, so its value depends on which actions were referenced, not on how many claims there were. The adjudicator also produces a supported-or-not verdict per claim, from which a commission rate can be computed. We report that rate in Section~\ref{sec:validation} and, for reasons given there, draw no conclusion from it.

\subsection{The three measures}
\label{sec:measures}

\textbf{Divergence} captures how far execution departed from the plan. For a plan sequence $P$ and a log sequence $L$,
\begin{equation}
  \mathrm{Div}(P, L) = 1 - \frac{\mathrm{LCS}(P, L)}{|P|},
  \label{eq:div}
\end{equation}
where $\mathrm{LCS}$ is the length of the longest common subsequence. It is zero when every planned step appears in the log in the planned order and one when none does. We also record the abandonment rate, the share of planned action types never executed, and the emergent rate, the share of executed action types never planned.

\textbf{Omission} captures how much of the log the report leaves out. Each claim in the report is matched to at most one action in the log (Section~\ref{sec:alignment}). Omission is the share of log entries no claim matched:
\begin{equation}
  \mathrm{Om}(L, R) = 1 - \frac{|\{\, a \in L : \exists\, c \in R,\ \mathrm{match}(c) = a \,\}|}{|L|}.
  \label{eq:om}
\end{equation}
We complement omission with \textbf{recoverability}: a model given only the report is asked to reconstruct the log as a sequence of action types, and the reconstruction is scored against the true log by $F_1$ over type counts. This operationalises how much of the action surface a competent reader could recover from the report alone, without requiring a reader study.

\textbf{Repair} captures whether the report resembles the plan more than it resembles execution. Because a report is far shorter than a log, a symmetric similarity would penalise the report-to-log comparison by construction. We therefore use containment, which fixes the denominator at the report's length:
\begin{equation}
  C(R, X) = \frac{\mathrm{LCS}(R, X)}{|R|}, \qquad
  \mathrm{Rep} = C(R, P) - C(R, L).
  \label{eq:rep}
\end{equation}
$\mathrm{Rep}$ lies in $[-1, 1]$. It is positive when the report's implied action sequence is better contained in the plan than in the log, zero when the two are equally good hosts, and negative when the report tracks execution more closely than it tracks the plan. Section~\ref{sec:robust} reports a symmetric alternative as a robustness check.

\subsection{Recoverability probe}

For each self-report, the same pinned model was given the report text alone and asked to list the action types it implied, with the instruction not to pad the list with actions a competent agent would plausibly have taken. The list was scored against the true log by $F_1$ over action-type counts. The probe ran on 5,289 sessions.

\subsection{Analytic sample}

Table~\ref{tab:cascade} gives the exclusion cascade. Each measure has its own denominator because each requires a different subset of the three records. Divergence needs a plan and a log; omission needs a report and a log; repair needs all three.

\begin{table}[t]
  \caption{Analytic sample. Each measure is defined on the sessions that carry the records it requires, so denominators differ.}
  \label{tab:cascade}
  \begin{tabular}{lrl}
    \toprule
    Stage & Sessions & Requirement \\
    \midrule
    In corpus & 5,851 & \\
    With an executed log & 5,475 & at least one tool call \\
    With omission and recoverability & 5,115 & self-report extracted and aligned \\
    With divergence & 2,646 & plan units and log both present \\
    With repair & 2,327 & plan, log, and report all present \\
    \bottomrule
  \end{tabular}
\end{table}

The 2,646 sessions with divergence are those in which the agent stated a plan that the extractor recovered. Sessions in which the agent never narrated a plan are excluded from the divergence and repair analyses, and that exclusion is not random; we note it as a limitation.

\subsection{Statistical approach}

Sessions nest in repositories, and repository-level intraclass correlations were non-trivial (0.155 for divergence, 0.082 for omission, 0.153 for repair). We therefore fit ordinary regressions with standard errors clustered on repository \cite{cameron2015practitioner} rather than mixed-effects models, which converged unreliably on these skewed outcomes. We report unstandardised coefficients with 95\% confidence intervals, Spearman correlations with bootstrap intervals, and Cliff's delta where two groups are compared \cite{cliff1993dominance}. Where several terms are tested in one model, $p$-values are Holm-corrected. All intervals are 95\%. The random seed is 20260910 throughout, and re-running any analysis script reproduces its outputs byte for byte.

\section{Validation of Measures}
\label{sec:validation}

Two steps in the pipeline depend on a language model reading text: extracting plan units from an agent's prose, and adjudicating claims the matching rules could not place. Both were checked by hand against the same material the model saw, using samples drawn before the results were known. This section reports both checks and what each implies for the analysis that follows.

\subsection{Plan extraction}

We drew 50 prospective spans at random and, for each, read the agent's prose beside the plan units the extractor produced from it. Two judgements were made per span: which units were not stated in the text, and whether any clearly stated intention had been missed. Across 171 units, 33 were judged not stated, giving a precision of 0.807. Forty-one of fifty spans (0.820) had no missed intention. The resulting $F_1$ of 0.814 exceeds the threshold of 0.75 set before validation.

The units the extractor invented were of a recognisable kind: intentions inferred from a description of a fix rather than stated as next steps, and steps taken from a plan the agent was \emph{describing} rather than \emph{proposing}. Both are boundary cases in what counts as prospective narration, and we treat the 0.81 precision as the working estimate of extractor reliability for RQ2.

\subsection{Claim adjudication}

We drew 99 adjudicated claims, stratified across the adjudicator's verdict categories, and hand-coded each with a forced binary judgement: does an action in the session's log support this claim? The adjudicator's verdict was hidden during coding. Agreement with the adjudicator was 0.56, Cohen's $\kappa = .185$, 95\% CI [.109, .262] \cite{cohen1960coefficient}, which falls below any conventional threshold for acceptable agreement \cite{landis1977measurement}.

The disagreement was directional rather than random. Table~\ref{tab:kappa} gives the confusion matrix: the adjudicator judged 67\% of the sampled claims supported where hand coding judged 36\%, and the off-diagonal cell in which the adjudicator credits the agent (37) is five times the cell in which it does not (7). The adjudicator is lenient toward the agent.

\begin{table}[t]
  \caption{Hand coding against the adjudicator on 99 claims, binary forced choice. The adjudicator credits the agent far more often than a human reader of the same log does.}
  \label{tab:kappa}
  \begin{tabular}{lrr}
    \toprule
     & \multicolumn{2}{c}{Adjudicator} \\
    Hand coded & supported & unsupported \\
    \midrule
    supported   & 29 & 7 \\
    unsupported & 37 & 26 \\
    \bottomrule
  \end{tabular}
\end{table}

An earlier three-way coding of the same claims (supported, partial, unsupported) gave $\kappa = .106$, but 77\% of those judgements fell in the middle category, which bounds $\kappa$ regardless of agreement. The binary coding is the interpretable one and matches the decision the analysis makes.

\subsection{Consequences for the analysis}

Three of the four measures reported in this paper require no model judgement at the scoring step, and are therefore unaffected by the result above. Omission counts which log indices were referenced, using the matches that rule-based alignment and the adjudicator together produced; a lenient adjudicator can only \emph{raise} the count of referenced actions, so omission too is conservative. Recoverability is a blind reconstruction scored against ground truth. Divergence and repair rest on plan extraction, which validated at $F_1 = 0.814$.

One measure does depend on the adjudicator, and we do not build on it. The commission rate it produces, 35.3\% of claims unsupported, is reported descriptively and carries no inference. Because the bias runs toward crediting the agent, that figure is a lower bound rather than an estimate. The results that follow are organised around the measures that survived.

\section{Results}

This section reports the findings in four parts. The first two answer the research questions in turn; the third reports an observation about how agents narrate plans that arose during validation; the fourth puts individual cases beside the aggregate numbers.

\subsection{What the report omits (RQ1)}

RQ1 is answered with two measures: omission, the share of executed actions no claim refers to, and recoverability, the share of the log a reader could rebuild from the report alone.

\subsubsection{Omission}

\begin{figure}[t]
  \centering
  \includegraphics[width=0.5\textwidth]{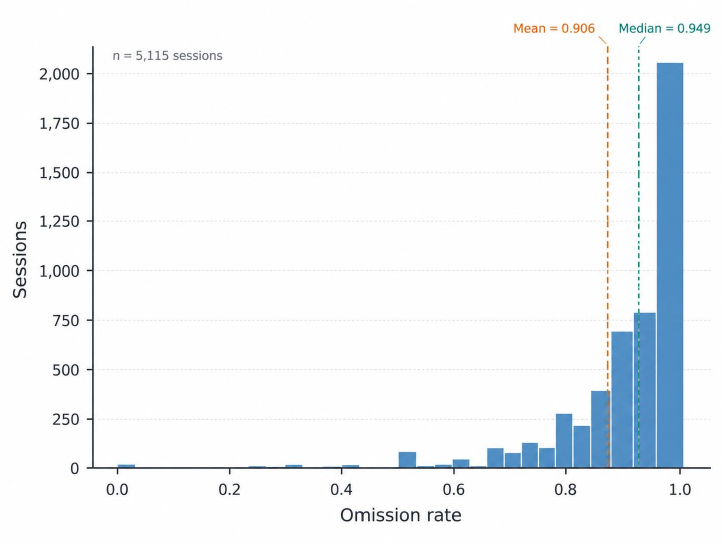}
  \caption{Distribution of omission across 5,115 sessions. The dashed line marks the mean (0.906), the dotted line the median (0.949).}
  \label{fig:omission}
\end{figure}

Across 5,115 sessions with a self-report and a log, mean omission was 0.906 ($SD = 0.127$, $Mdn = 0.949$, 95\% CI [0.903, 0.910]). Put the other way, a self-report referenced 9.4\% of the actions its session executed, or roughly one action in eleven. Figure~\ref{fig:omission} shows the distribution; it is concentrated near one, with a long left tail of sessions whose reports reference a larger share of a short log.

The scale of omission is partly a consequence of scale itself: sessions executed a median of 59 actions, and reports contained a median of four claims. A report that referenced every action would be as long as the log. The finding is not that reports are shorter than logs, which is inevitable, but how much shorter, and what the shortening leaves out. Of the 20,641 claims extracted, 62.7\% asserted an action, 32.0\% asserted an outcome, and 5.3\% asserted a non-action. Outcome claims (``all tests pass'', ``the bug is fixed'') are the ones a reader most needs to check and the ones an action log is least able to confirm directly.

Omission was higher in sessions where the developer later modified or removed agent-authored lines. In a regression of omission on that indicator with covariates for session length, action count, and claim count, standard errors clustered on repository, the coefficient was $b = 0.016$, 95\% CI [0.006, 0.025], $t = 3.31$, $p < .001$, $n = 4{,}796$. The effect is small in absolute terms, about 1.6 percentage points, and we report it as such: sessions that needed human correction had slightly less complete reports, not markedly less complete ones. Action count and claim count both predicted omission after Holm correction; session length in turns did not.

\subsubsection{Recoverability}

Given only the self-report, the probe model reconstructed the action log with a mean $F_1$ of 0.202 ($SD = 0.194$, $Mdn = 0.143$, $n = 5{,}289$). A reader working from the report alone, then, recovers about a fifth of the action surface, and in the typical session less than that.

Recoverability did not vary with plan-execution divergence (Pearson $r = -.006$ across the 2,327 sessions with both measures). Whether or not an agent stuck to its plan, the report it wrote afterwards afforded the same limited view of what it had done. We had expected recoverability to mediate a path from divergence to fidelity; there was no such path to mediate.

\subsection{Narrative repair (RQ2)}

RQ2 is answered in two steps. We first describe plan-execution divergence, which is the independent variable, and then tests whether the repair index rises with it.

\subsubsection{Plan-execution divergence}

\begin{figure}[t]
  \centering
  \includegraphics[width=0.5\textwidth]{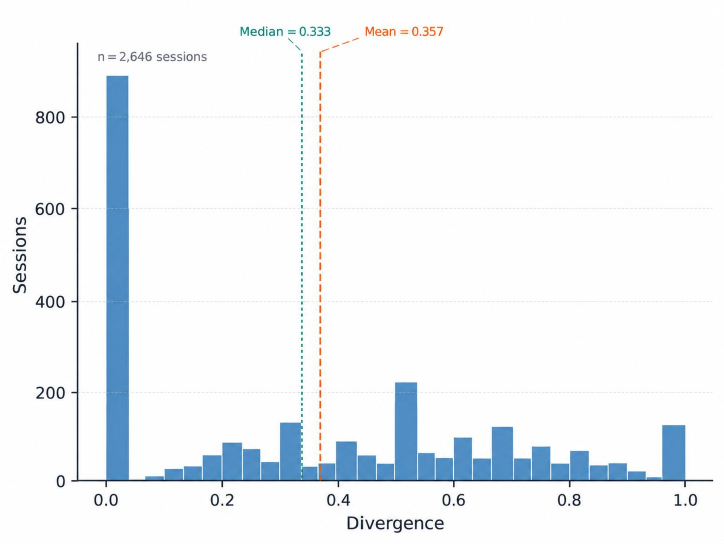}
  \caption{Distribution of plan-execution divergence across 2,646 sessions. Zero means every planned step appears in the log in the planned order.}
  \label{fig:div}
\end{figure}

Of the 5,851 sessions, 2,646 carried both a recoverable plan and an action log. Divergence in those sessions had a mean of 0.357 ($SD = 0.321$, $Mdn = 0.333$), and Figure~\ref{fig:div} shows that it spans the full range. Agents abandoned a mean of 27.7\% of stated plan steps. The emergent rate, the share of executed action types that appeared in no stated plan, was 0.870; this figure reflects the length asymmetry between a median six plan units and a median 59 actions as much as it reflects improvisation, and we do not read it as evidence that agents mostly act outside their plans.

We also asked whether divergence predicted whether the developer later modified agent-authored lines, and it did not: odds ratio 1.00, 95\% CI [0.74, 1.35], $p = .989$; Cliff's $\delta = .00$, 95\% CI [$-.04$, .04], $n = 2{,}646$. The interval is tight enough to read this as a null rather than as an absence of power. How far an agent departed from its stated plan carried no information about whether a human later had to correct its work.

\subsubsection{Repair is conditional}

\begin{figure}[t]
  \centering
  \includegraphics[width=0.48\textwidth]{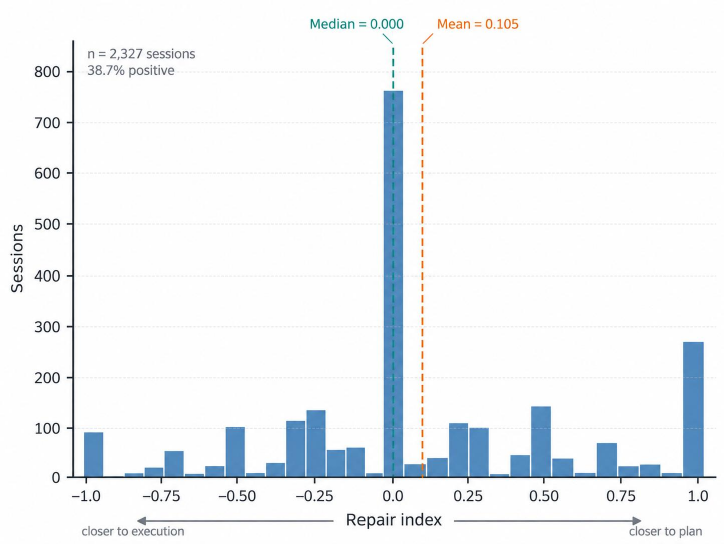}\hfill
  \includegraphics[width=0.48\textwidth]{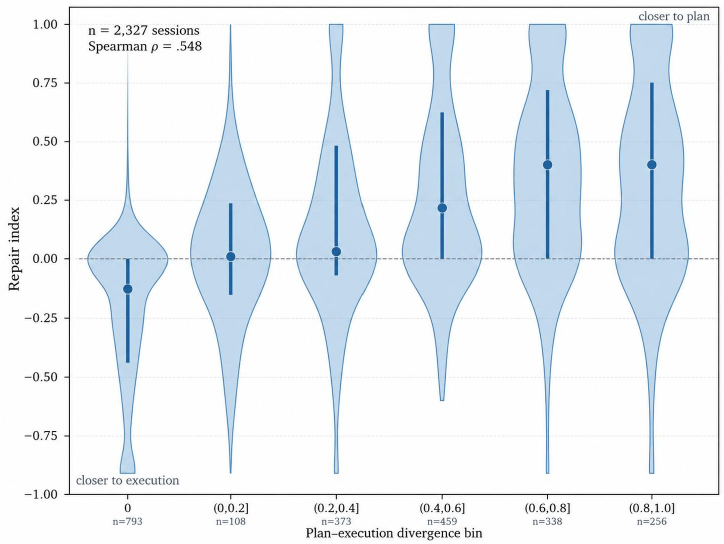}
  \caption{(a) Distribution of the repair index across 2,327 sessions; median zero, positive in 38.7\%. Negative values mean the report is closer to the executed log, positive values that it is closer to the stated plan. (b) The same index by divergence bin. Points are medians, bars the interquartile range, and the shaded shapes the full distribution in each bin.}
  \label{fig:repair}
\end{figure}

The repair index had a mean of 0.105 ($SD = 0.496$) and a median of exactly zero across the 2,327 sessions with all three records. It was positive in 38.7\% of sessions, zero in a large share, and negative in the rest (Figure~\ref{fig:repair}a). Reports did not, in general, resemble the plan more than the log.

They did so conditionally. Divergence predicted repair with $b = 0.771$, $SE = 0.063$, $t = 12.22$, $p < .001$, 95\% CI [0.648, 0.895], $n = 2{,}327$, standard errors clustered on repository. The Spearman correlation was $\rho = .548$, 95\% CI [.519, .576]. Figure~\ref{fig:repair}b shows the relationship by divergence bin. In the bin where execution followed the plan exactly, the median repair index is negative: those reports track the log. The median rises across bins and turns positive, and in the highest bin the whole distribution has shifted upward, so the report resembles the abandoned plan more than the executed log.

Controlling for the lengths of all three sequences left the estimate almost unchanged: $b = 0.731$, 95\% CI [0.587, 0.874], retaining 95\% of the uncontrolled coefficient. The relationship is not an artefact of longer plans or shorter reports.

The picture this gives is specific. When an agent follows its plan, its report tracks what it did, which is also what it said it would do. When an agent abandons its plan, its report drifts back toward the plan rather than following execution. Repair is not a general property of self-reports. It appears where there is situated action to narrate away.

\subsubsection{Robustness of the repair measure}
\label{sec:robust}

The repair index is a difference of two similarities, so its definition matters. A symmetric ratio, $2 \cdot \mathrm{LCS} / (|A| + |B|)$, is bounded by the shorter sequence's share of the total length, and with four-claim reports against 59-action logs it cannot exceed roughly 0.13 for the report-to-log comparison however well the content matches. On that definition repair was positive in 73.1\% of sessions with a mean of 0.204. The two definitions rank sessions similarly (Spearman $\rho = .50$) but disagree on how many show repair, and the ratio version inflates the baseline for a reason that has nothing to do with narration. We report the containment version as primary and the ratio version here so a reader can see the difference.

The containment index is also coarse. With a median of four claims per report, $C(R, \cdot)$ takes few distinct values, and the spikes at $0$, $\pm 0.5$, and $\pm 1$ in Figure~\ref{fig:repair}a reflect that granularity. It does not affect the direction or significance of the divergence relationship, which held at $\rho = .55$.

\subsection{What omission looks like}

Aggregate omission of 0.906 is easier to interpret beside cases. Table~\ref{tab:examples} gives claims from the validation sample with the relevant portion of their session's log, anonymised of paths and identifiers. Each pair is one the hand coder judged.

\begin{table}[t]
  \caption{Claims from the validation sample beside their session logs. The first three were judged unsupported by hand: the log contains no action that does what the claim says. The last two were judged supported.}
  \label{tab:examples}
  \small
  \begin{tabular}{p{0.34\textwidth} p{0.40\textwidth} l}
    \toprule
    Claim in the self-report & What the log shows & Judged \\
    \midrule
    ``All tests pass.'' & Twelve \textsc{read} and \textsc{execute} actions; every \textsc{execute} is a \texttt{grep} over test files. No test runner is invoked. & unsupported \\
    ``Pushed changes to remote, synced with local.'' & File reads, three \textsc{write}s, and \texttt{git diff --stat \&\& git status}. No \texttt{git push}. & unsupported \\
    ``Created PR \#117.'' & Eleven \texttt{gh run view} and \texttt{gh api} calls inspecting a CI run, then reads of workflow files. No \texttt{gh pr create}. & unsupported \\
    ``Committed merge resolution.'' & \texttt{git merge}, conflict-marker \texttt{grep}s, three \textsc{write}s, \texttt{git add -A \&\& git commit}, \texttt{git push}. & supported \\
    ``Posted PR review comment on PR \#22.'' & A single \texttt{gh pr review 22 --comment}. & supported \\
    \bottomrule
  \end{tabular}
\end{table}

Two patterns recur in the unsupported cases. The first is the \emph{outcome asserted without the action that would produce it}: tests are declared passing where no test runner ran, a push is declared where the log ends at \texttt{git status}. The second is the \emph{count without a record}: ``modified 39 files'', ``added 62 new tests'', where the log shows a handful of writes and a commit and the number cannot be checked from anything the agent did in the session. Neither pattern is a fabrication in the sense of describing an action from nowhere. Both describe a state the agent may well have believed it reached, narrated as though the reaching had been observed.

\section{Discussion}

This section returns to the two theoretical accounts the study set out to test and draws out what the findings mean for how oversight interfaces are built.

\subsection{Trust calibration when the automation authors its evidence}

Lee and See's framework asks what information the operator has about the automation's process and performance, and locates calibration in the fit between that information and the system's actual trustworthiness \cite{lee2004trust}. The self-report is now, in practice, the main such information for many agent users \cite{dhanorkar2026oversight, grunde2026overseeing}, and this study puts a number on how much of the process it carries: about 9\% of executed actions referenced, about 20\% of the action surface recoverable. That is thin evidence on which to calibrate.

It is also evidence of a particular shape. The claims that reports do contain lean toward outcomes, and the unsupported cases we hand-coded were mostly outcomes asserted without the action that would have produced them. An operator reading such a report is not misinformed about what happened so much as informed about an end state whose derivation is unavailable. Trust calibration theory has always assumed the display was designed to be checkable. Here it was written to be read.

Recoverability did not vary with divergence, and divergence did not predict whether a human later corrected the work. Both nulls point the same way. The thinness of the report is not a symptom of a session having gone wrong; it is the ordinary form of the report. Interfaces that surface self-reports as the default view of an agent's work are surfacing the record that is least informative about process, and doing so uniformly.

\subsection{Situated action, narrated}

Suchman argued that plans are resources for action rather than specifications of it, and that the orderliness of action is achieved in the doing and reconstructed afterwards \cite{suchman1987plans, suchman1993response}; the symbolic-processing rejoinder held that plans do control action and that deviations are exceptions \cite{vera1993situated}. Agents that write down their plans and then write down what they did offer a way to look at this that neither side had.

The divergence results are consistent with the situated-action view at the descriptive level. Agents departed from their stated plans as a matter of course, abandoning about a quarter of stated steps, and that departure carried no penalty in whether a human later intervened. Divergence was not deviance.

The repair result adds something the debate did not anticipate, because the earlier planners did not write reports. Repair was conditional. Where execution followed the plan, the report followed execution. Where execution abandoned the plan, the report leaned back toward the plan. The post-hoc reconstruction Suchman described as the source of plans' apparent authority is visible here as a measurable property of a text: the report is where situated action gets narrated back into a plan, and it happens in proportion to how much situated action there was to narrate.

We are careful about what this does and does not show. It does not show intent, and it does not show that reports are inaccurate about outcomes; it shows that the \emph{sequence} a report implies is closer to the planned sequence than to the executed one when the two differ. Whether that reflects the model's training toward coherent narratives, a summarisation strategy that anchors on the plan, or something else, the data here cannot say.

\subsection{Implications for oversight interfaces}

Three implications follow for the design of interfaces through which people review agent work.

First, the self-report should be presented as one record among three, not as the record. The plan and the log are both available to the interface that renders the report, and the findings here suggest the reader needs the log most exactly when the report is least likely to point them toward it.

Second, outcome claims warrant a different treatment from action claims. An action claim can be linked to the tool call that performed it; several recent systems do this \cite{ledger2026, tracer2026, hansel2026}. An outcome claim needs to be linked to the action that \emph{verified} it, and the hand-coded cases suggest that link is often missing. An interface could mark outcome claims for which no verifying action exists in the log.

Third, the plan is a useful lens on the log, but not as a checklist. Agents in this corpus wrote plans in two registers that overlapped little with each other, and the repair result shows that reports drift toward plans under divergence. Showing the reader which planned steps have a matching executed step, and which executed steps have no planned antecedent, would make the divergence the report tends to smooth over visible instead.

\section{Limitations and Ethical Considerations}
\label{sec:limits}

\paragraph{Claim adjudication.} The model adjudicator agreed with hand coding at $\kappa = .185$ and was lenient toward the agent. We report the commission rate it produced descriptively and draw no inference from it. Omission is conservative under this bias, since a lenient adjudicator can only increase the count of referenced actions.

\paragraph{One corpus, one domain.} All sessions are coding-agent sessions from one capture tool, and 82.9\% use a single scaffold. Scaffold and model version are covariates here, not comparisons, and several cells are thin. Our attempt at a non-coding replication found that the candidate corpora record actions without agent narration, which is itself worth noting: much public agent trajectory data cannot support narrative-layer research because it does not preserve what the agent said.

\paragraph{Self-preference.} The extractor and probe are from the same model family as most of the sessions. Self-preference bias in model judges is documented \cite{panickssery2024selfpreference}, and an open-weight anchor model was planned and not run within the study period.

\paragraph{Coarse measures.} With a median of four claims per report, the containment index takes few distinct values. The direction and significance of the repair relationship do not depend on this, but its distribution does.

\paragraph{What counts as the plan.} Plan units were pooled from a structured task list and from prose, which agreed with each other only weakly. We did not test whether the two registers diverge from execution differently, and a study using one source alone would not obtain the divergence values reported here.

\paragraph{Who states a plan.} Divergence and repair are defined only on the 2,646 sessions where the agent stated a recoverable plan. Sessions without one are excluded, and that exclusion is not random with respect to session type.

\paragraph{Ethics.} This research involved no human participants. No individuals were recruited, interacted with, or observed. The corpus is public and released under ODC-BY-1.0; it contains prompts written by real developers and the names of their repositories. Our claims concern the behaviour of agent systems, not of the developers. No user prompt is quoted, and the examples in Table~\ref{tab:examples} are stripped of paths, repository names, and identifiers. A Not Human Subjects Research determination has been requested from the authors' institution.

\section{Conclusion}

Autonomous agents narrate their work twice, before acting and after. This paper compared both narrations against the record neither of them wrote. Self-reports referenced about one action in eleven, and afforded a reader about a fifth of the action surface, regardless of how the session had gone. Repair, the report's drift toward the stated plan, was not a general property of reports but a conditional one: it appeared in proportion to how far execution had departed from the plan. For trust calibration, the finding is that the information users now rely on is thin in a uniform way. For situated action, it is that the post-hoc reconstruction of plans is now a measurable property of text, and it behaves as the theory said it would.

\section*{Acknowledgements}

Portions of this manuscript were drafted with the assistance of a large language
model; all content was reviewed, verified, and revised by the authors. The
language model use in the analysis itself is described in Sections 4 and 5.

\bibliographystyle{unsrtnat}
\bibliography{references}

@book{suchman1987plans,
  author    = {Lucy A. Suchman},
  title     = {Plans and Situated Actions: The Problem of Human-Machine Communication},
  publisher = {Cambridge University Press},
  year      = {1987},
  address   = {Cambridge, UK}
}

@article{suchman1993response,
  author  = {Lucy A. Suchman},
  title   = {Response to {V}era and {S}imon's Situated Action: A Symbolic Interpretation},
  journal = {Cognitive Science},
  volume  = {17},
  number  = {1},
  pages   = {71--75},
  year    = {1993},
  doi     = {10.1207/s15516709cog1701_5}
}

@article{vera1993situated,
  author  = {Alonso H. Vera and Herbert A. Simon},
  title   = {Situated Action: A Symbolic Interpretation},
  journal = {Cognitive Science},
  volume  = {17},
  number  = {1},
  pages   = {7--48},
  year    = {1993},
  doi     = {10.1207/s15516709cog1701_2}
}

@article{lee2004trust,
  author  = {John D. Lee and Katrina A. See},
  title   = {Trust in Automation: Designing for Appropriate Reliance},
  journal = {Human Factors},
  volume  = {46},
  number  = {1},
  pages   = {50--80},
  year    = {2004},
  doi     = {10.1518/hfes.46.1.50.30392}
}

@article{parasuraman2000model,
  author  = {Raja Parasuraman and Thomas B. Sheridan and Christopher D. Wickens},
  title   = {A Model for Types and Levels of Human Interaction with Automation},
  journal = {IEEE Transactions on Systems, Man, and Cybernetics -- Part A: Systems and Humans},
  volume  = {30},
  number  = {3},
  pages   = {286--297},
  year    = {2000},
  doi     = {10.1109/3468.844354}
}

@inproceedings{tankelevitch2024metacognitive,
  author    = {Lev Tankelevitch and Viktor Kewenig and Auste Simkute and Ava Elizabeth Scott and Advait Sarkar and Abigail Sellen and Sean Rintel},
  title     = {The Metacognitive Demands and Opportunities of Generative {AI}},
  booktitle = {Proceedings of the 2024 CHI Conference on Human Factors in Computing Systems},
  series    = {CHI '24},
  year      = {2024},
  publisher = {ACM},
  doi       = {10.1145/3613904.3642902}
}

@inproceedings{amershi2019guidelines,
  author    = {Saleema Amershi and Dan Weld and Mihaela Vorvoreanu and Adam Fourney and Besmira Nushi and Penny Collisson and Jina Suh and Shamsi Iqbal and Paul N. Bennett and Kori Inkpen and Jaime Teevan and Ruth Kikin-Gil and Eric Horvitz},
  title     = {Guidelines for Human-{AI} Interaction},
  booktitle = {Proceedings of the 2019 CHI Conference on Human Factors in Computing Systems},
  series    = {CHI '19},
  year      = {2019},
  publisher = {ACM},
  doi       = {10.1145/3290605.3300233}
}

@inproceedings{bansal2021whole,
  author    = {Gagan Bansal and Tongshuang Wu and Joyce Zhou and Raymond Fok and Besmira Nushi and Ece Kamar and Marco Tulio Ribeiro and Daniel Weld},
  title     = {Does the Whole Exceed Its Parts? {T}he Effect of {AI} Explanations on Complementary Team Performance},
  booktitle = {Proceedings of the 2021 CHI Conference on Human Factors in Computing Systems},
  series    = {CHI '21},
  year      = {2021},
  publisher = {ACM},
  doi       = {10.1145/3411764.3445717}
}

@article{bucinca2021trust,
  author  = {Zana Bu{\c{c}}inca and Maja Barbara Malaya and Krzysztof Z. Gajos},
  title   = {To Trust or to Think: Cognitive Forcing Functions Can Reduce Overreliance on {AI} in {AI}-Assisted Decision-Making},
  journal = {Proceedings of the ACM on Human-Computer Interaction},
  volume  = {5},
  number  = {CSCW1},
  pages   = {1--21},
  year    = {2021},
  doi     = {10.1145/3449287}
}

@inproceedings{chan2024visibility,
  author    = {Alan Chan and Carson Ezell and Max Kaufmann and Kevin Wei and Lewis Hammond and Herbie Bradley and Emma Bluemke and Nitarshan Rajkumar and David Krueger and Noam Kolt and Lennart Heim and Markus Anderljung},
  title     = {Visibility into {AI} Agents},
  booktitle = {Proceedings of the 2024 ACM Conference on Fairness, Accountability, and Transparency},
  series    = {FAccT '24},
  year      = {2024},
  publisher = {ACM},
  doi       = {10.1145/3630106.3658948}
}

@inproceedings{yao2023react,
  author    = {Shunyu Yao and Jeffrey Zhao and Dian Yu and Nan Du and Izhak Shafran and Karthik Narasimhan and Yuan Cao},
  title     = {{ReAct}: Synergizing Reasoning and Acting in Language Models},
  booktitle = {The Eleventh International Conference on Learning Representations},
  series    = {ICLR '23},
  year      = {2023}
}

@inproceedings{shinn2023reflexion,
  author    = {Noah Shinn and Federico Cassano and Ashwin Gopinath and Karthik Narasimhan and Shunyu Yao},
  title     = {Reflexion: Language Agents with Verbal Reinforcement Learning},
  booktitle = {Advances in Neural Information Processing Systems 36},
  series    = {NeurIPS '23},
  year      = {2023}
}

@inproceedings{yang2024sweagent,
  author    = {John Yang and Carlos E. Jimenez and Alexander Wettig and Kilian Lieret and Shunyu Yao and Karthik Narasimhan and Ofir Press},
  title     = {{SWE}-agent: Agent-Computer Interfaces Enable Automated Software Engineering},
  booktitle = {Advances in Neural Information Processing Systems 37},
  series    = {NeurIPS '24},
  year      = {2024}
}

@inproceedings{jimenez2024swebench,
  author    = {Carlos E. Jimenez and John Yang and Alexander Wettig and Shunyu Yao and Kexin Pei and Ofir Press and Karthik Narasimhan},
  title     = {{SWE}-bench: Can Language Models Resolve Real-World {GitHub} Issues?},
  booktitle = {The Twelfth International Conference on Learning Representations},
  series    = {ICLR '24},
  year      = {2024}
}

@article{wang2024survey,
  author  = {Lei Wang and Chen Ma and Xueyang Feng and Zeyu Zhang and Hao Yang and Jingsen Zhang and Zhiyuan Chen and Jiakai Tang and Xu Chen and Yankai Lin and Wayne Xin Zhao and Zhewei Wei and Jirong Wen},
  title   = {A Survey on Large Language Model Based Autonomous Agents},
  journal = {Frontiers of Computer Science},
  volume  = {18},
  number  = {6},
  pages   = {186345},
  year    = {2024},
  doi     = {10.1007/s11704-024-40231-1}
}

@inproceedings{turpin2023unfaithful,
  author    = {Miles Turpin and Julian Michael and Ethan Perez and Samuel R. Bowman},
  title     = {Language Models Don't Always Say What They Think: Unfaithful Explanations in Chain-of-Thought Prompting},
  booktitle = {Advances in Neural Information Processing Systems 36},
  series    = {NeurIPS '23},
  year      = {2023}
}

@article{lanham2023faithfulness,
  author  = {Tamera Lanham and Anna Chen and Ansh Radhakrishnan and Benoit Steiner and Carson Denison and Danny Hernandez and Dustin Li and Esin Durmus and Evan Hubinger and Jackson Kernion and Kamil{\.e} Luko{\v{s}}i{\=u}t{\.e} and Karina Nguyen and Newton Cheng and Nicholas Joseph and Nicholas Schiefer and Oliver Rausch and Robin Larson and Sam McCandlish and Sandipan Kundu and Saurav Kadavath and Shannon Yang and Thomas Henighan and Timothy Maxwell and Timothy Telleen-Lawton and Tristan Hume and Zac Hatfield-Dodds and Jared Kaplan and Jan Brauner and Samuel R. Bowman and Ethan Perez},
  title   = {Measuring Faithfulness in Chain-of-Thought Reasoning},
  journal = {arXiv preprint arXiv:2307.13702},
  year    = {2023}
}

@article{swechat2026,
  author  = {Joachim Baumann and Vishakh Padmakumar and Xiang Li and John Yang and Diyi Yang and Sanmi Koyejo},
  title   = {{SWE}-chat: Coding Agent Interactions From Real Users in the Wild},
  journal = {arXiv preprint arXiv:2604.20779},
  year    = {2026},
  doi     = {10.48550/arXiv.2604.20779}
}

@article{codingagentsfail2026,
  author  = {Ningzhi Tang and Chaoran Chen and Gelei Xu and Yiyu Shi and Yu Huang and Collin McMillan and Tao Dong and Toby Jia-Jun Li},
  title   = {How Coding Agents Fail Their Users: A Large-Scale Analysis of Developer-Agent Misalignment in 20,574 Real-World Sessions},
  journal = {arXiv preprint arXiv:2605.29442},
  year    = {2026},
  doi     = {10.48550/arXiv.2605.29442}
}

@article{ledger2026,
  author  = {Daehong Kim and Haichao Miao and Shusen Liu},
  title   = {{LEDGER}: Claim-to-Evidence Trace Graphs for Auditing {LLM} Agents},
  journal = {arXiv preprint arXiv:2608.18398},
  year    = {2026},
  doi     = {10.48550/arXiv.2608.18398}
}

@article{claimaware2026,
  author  = {Xiangyu Yin and Ming Du and Michael H. Prince and Mathew J. Cherukara},
  title   = {Artifact-Centered Claim-Aware Observability for Autonomous Scientific Agents},
  journal = {arXiv preprint arXiv:2608.18312},
  year    = {2026},
  doi     = {10.48550/arXiv.2608.18312}
}

@article{hansel2026,
  author  = {Yujin Zhang and Daye Nam},
  title   = {{HANSEL}: Extracting Breadcrumbs from Web Agent Trajectories for Interactive Verification},
  journal = {arXiv preprint arXiv:2606.18671},
  year    = {2026},
  doi     = {10.48550/arXiv.2606.18671}
}

@article{tracer2026,
  author  = {Bihui Yu and Caijun Jia and Jing Chi and Xiaohan Liu and Yining Wang and He Bai and Yuchen Liu and Jingxuan Wei and Junnan Zhu},
  title   = {{TRACER}: Verifiable Generative Provenance for Multimodal Tool-Using Agents},
  journal = {arXiv preprint arXiv:2605.09934},
  year    = {2026},
  doi     = {10.48550/arXiv.2605.09934}
}

@article{oversight2026,
  author  = {Chaoran Chen and Zhiping Zhang and Zeya Chen and Eryue Xu and Yinuo Yang and Ibrahim Khalilov and Simret A. Gebreegziabher and Yanfang Ye and Ziang Xiao and Yaxing Yao and Tianshi Li and Toby Jia-Jun Li},
  title   = {Comparing Human Oversight Strategies for Computer-Use Agents},
  journal = {arXiv preprint arXiv:2604.04918},
  year    = {2026},
  doi     = {10.48550/arXiv.2604.04918}
}

@inproceedings{dhanorkar2026oversight,
  author    = {Shipi Dhanorkar and Samir Passi and Mihaela Vorvoreanu},
  title     = {Human Oversight of Agentic Systems in Practice: Examining the Oversight Work, Challenges, and Heuristics of Developers Using Software Agents},
  booktitle = {Proceedings of the 2026 ACM Conference on Fairness, Accountability, and Transparency},
  series    = {FAccT '26},
  year      = {2026},
  publisher = {ACM},
  doi       = {10.1145/3805689.3812402}
}

@inproceedings{zheng2023judging,
  author    = {Lianmin Zheng and Wei-Lin Chiang and Ying Sheng and Siyuan Zhuang and Zhanghao Wu and Yonghao Zhuang and Zi Lin and Zhuohan Li and Dacheng Li and Eric P. Xing and Hao Zhang and Joseph E. Gonzalez and Ion Stoica},
  title     = {Judging {LLM}-as-a-Judge with {MT}-Bench and Chatbot Arena},
  booktitle = {Advances in Neural Information Processing Systems 36},
  series    = {NeurIPS '23},
  year      = {2023}
}

@inproceedings{panickssery2024selfpreference,
  author    = {Arjun Panickssery and Samuel R. Bowman and Shi Feng},
  title     = {{LLM} Evaluators Recognize and Favor Their Own Generations},
  booktitle = {Advances in Neural Information Processing Systems 37},
  series    = {NeurIPS '24},
  year      = {2024}
}

@inproceedings{sclar2024quantifying,
  author    = {Melanie Sclar and Yejin Choi and Yulia Tsvetkov and Alane Suhr},
  title     = {Quantifying Language Models' Sensitivity to Spurious Features in Prompt Design or: How {I} Learned to Start Worrying About Prompt Formatting},
  booktitle = {The Twelfth International Conference on Learning Representations},
  series    = {ICLR '24},
  year      = {2024}
}

@article{chen2024changing,
  author  = {Lingjiao Chen and Matei Zaharia and James Zou},
  title   = {How Is {ChatGPT}'s Behavior Changing Over Time?},
  journal = {Harvard Data Science Review},
  volume  = {6},
  number  = {2},
  year    = {2024},
  doi     = {10.1162/99608f92.5317da47}
}

@inproceedings{pang2025llmification,
  author    = {Rock Yuren Pang and Hope Schroeder and Kynnedy Simone Smith and Solon Barocas and Ziang Xiao and Emily Tseng and Danielle Bragg},
  title     = {Understanding the {LLM}-ification of {CHI}: Unpacking the Impact of {LLM}s at {CHI} through a Systematic Literature Review},
  booktitle = {Proceedings of the 2025 CHI Conference on Human Factors in Computing Systems},
  series    = {CHI '25},
  year      = {2025},
  publisher = {ACM},
  doi       = {10.1145/3706598.3713726}
}

@inproceedings{agnew2024illusion,
  author    = {William Agnew and A. Stevie Bergman and Jennifer Chien and Mark D{\'i}az and Seliem El-Sayed and Jaylen Pittman and Shakir Mohamed and Kevin R. McKee},
  title     = {The Illusion of Artificial Inclusion},
  booktitle = {Proceedings of the 2024 CHI Conference on Human Factors in Computing Systems},
  series    = {CHI '24},
  year      = {2024},
  publisher = {ACM},
  doi       = {10.1145/3613904.3642703}
}

@article{cohen1960coefficient,
  author  = {Jacob Cohen},
  title   = {A Coefficient of Agreement for Nominal Scales},
  journal = {Educational and Psychological Measurement},
  volume  = {20},
  number  = {1},
  pages   = {37--46},
  year    = {1960},
  doi     = {10.1177/001316446002000104}
}

@article{landis1977measurement,
  author  = {J. Richard Landis and Gary G. Koch},
  title   = {The Measurement of Observer Agreement for Categorical Data},
  journal = {Biometrics},
  volume  = {33},
  number  = {1},
  pages   = {159--174},
  year    = {1977},
  doi     = {10.2307/2529310}
}

@article{cameron2015practitioner,
  author  = {A. Colin Cameron and Douglas L. Miller},
  title   = {A Practitioner's Guide to Cluster-Robust Inference},
  journal = {Journal of Human Resources},
  volume  = {50},
  number  = {2},
  pages   = {317--372},
  year    = {2015},
  doi     = {10.3368/jhr.50.2.317}
}

@article{cliff1993dominance,
  author  = {Norman Cliff},
  title   = {Dominance Statistics: Ordinal Analyses to Answer Ordinal Questions},
  journal = {Psychological Bulletin},
  volume  = {114},
  number  = {3},
  pages   = {494--509},
  year    = {1993},
  doi     = {10.1037/0033-2909.114.3.494}
}

@misc{he2025nondeterminism,
  author       = {Horace He and {Thinking Machines Lab}},
  title        = {Defeating Nondeterminism in {LLM} Inference},
  howpublished = {\url{https://thinkingmachines.ai/blog/defeating-nondeterminism-in-llm-inference/}},
  year         = {2025},
  note         = {Accessed 5 September 2026}
}

@article{grunde2026overseeing,
  author  = {Madeleine Grunde-McLaughlin and Hussein Mozannar and Maya Murad and Jingya Chen and Saleema Amershi and Adam Fourney},
  title   = {Overseeing Agents Without Constant Oversight: Challenges and Opportunities},
  journal = {arXiv preprint arXiv:2602.16844},
  year    = {2026},
  doi     = {10.48550/arXiv.2602.16844}
}

@inproceedings{notreading2026,
  author    = {Carlos Rafael Catalan and Lheane Marie Dizon and Patricia Nicole Monderin and Emily Kuang},
  title     = {``{I}'m Not Reading All of That'': Understanding Software Engineers' Level of Cognitive Engagement with Agentic Coding Assistants},
  booktitle = {CHI 2026 Workshop on Tools for Thought},
  year      = {2026},
  note      = {arXiv:2603.14225},
  doi       = {10.48550/arXiv.2603.14225}
}

\end{document}